\documentclass[a4paper, 12pt]{article}
\usepackage{amsmath, amssymb, graphics}
\newcommand{\mathsym}[1]{{}}
\newcommand{\unicode}[1]{{}}

\catcode`\@=11

\@addtoreset{equation}{section} %\catcode`\@=10

\def\dddot#1{\mathinner{\buildrel\vbox{\kern5pt\hbox{...}}\over{#1}}}
\def\ddddot#1{\mathinner{\buildrel\vbox{\kern5pt\hbox{....}}\over{#1}}}

\begin{document}

\begin {center}
{\Large Algebraic Properties for Certain Form of the Members of Sequence on Generalized Modified Camassa-Holm Equation}\\[3 mm]
{\small K Krishnakumar$ ^ {1}$ , A Durga Devi $ ^{2*}$,  V Srinivasan$ ^ 1$ and PGL Leach$ ^ 3$\\[3 mm]
$^ 1 $ Department of Mathematics, Srinivasa Ramanujan Centre, SASTRA Deemed to be University, Kumbakonam 612 001, India.\\$^ 2 $ Department of Physics, Srinivasa Ramanujan Centre, SASTRA Deemed to be University, Kumbakonam 612 001, India.\\$^3 $Department of Mathematics, Durban University of Technology, PO Box 1334,\\ Durban ~4000,~ Republic of South Africa.}
\end{center}
\begin {abstract}
We study the symmetry and integrability of a Generalized Modified Camassa-Holm Equation (GMCH) of the form  $$u_{t}-u_{xxt}+2nu_{x}(u^2-u_{x}^2)^{n-1}(u-u_{xx})^2+(u^2-u_{x}^2)^{n}(u_{x}-u_{xxx})=0.$$ We observe that for increasing values of $n\in \mathbb{N}$, $\mathbb{N}$ denotes natural number, the above equation gives a family of equations in which nonlinearity is rapidly increasing as $n$ increases. However, this family has similar form of symmetries except the values of $n$. Interestingly the resultant second-order nonlinear ODE which is to be obtained from GMCH equation has eight dimensional symmetries. Hence the second-order nonlinear ODE is linearizable. Finally we conclude that the resultant second-order nonlinear ordinary differential equation which is obtained from the family of GMCH passes the Painlev\'e Test also it posses the similar form of leading order, resonances and truncated series solution too.
\end {abstract}

\section {Introduction}
Researchers who are working in such a field of Mathematical Physics, Engineering, Mathematical Biology etc. are developing mathematical models to understand the behavior of the physical phenomena which are arising in their respective fields.  Especially they are converting their physical phenomena into system of differential equations. The deterministic mathematical descriptions of such a differential equations can be cast into appropriate mathematical models through nonlinear ordinary differential equations, partial differential equations and so on. Among them partial differential equations have been mostly used to study the nature of the systems. For example in $1834$ Korteweg–de Vries (KdV) equation which is nonlinear partial differential equation studied experimentally to analyze waves on shallow water surfaces by John Scott Russell then Lord Rayleigh and Joseph Boussinesq studied theoretical investigation about this equation by around $1870$ and followed by Korteweg and De Vries in $1895$. Similarly, there are many wave equations have been investigated in various fields which is quietly related to mathematics. Few of them are Korteweg–de Vries (KdV) Equation, nonlinear coupled KdV Equation \cite{Korteweg 95 a}, Camassa-Holm (CH) Equation \cite{ Camassaholm}, Modified Camassa-Holm (MCH) Equation \cite{Majeed, ConstantinLannes}. Numerous methods have been discussed to solve above mentioned equations \cite{Hirota 71 a, Hirota 73 a, Hirota 73 b, Hirota 76 a, Hirota 74 a, Gui-2013, Fuch-1996, Fokas1995, Xia-2018, Fu-2013}.

\strut\hfill

In general, finding a general non-trivial solutions of a nonlinear differential equation is very hard to achieve in most instances. However, Lie symmetry has been used as an important tool that provide a systematic method to find the solution of the given differential equations successfully \cite{Lie74a, Olver-86, Leach 88 a, Leach 99, Leach 12 a, Andriopoulos09a, Krishnakumar 14 a, Krishnakumar 14 b}. By using the symmetries, one able to get algebraic equations from the given ordinary or partial differential equations through performing the reduction of order the differential equations \cite{Krishnakumar 14 c}. Hence, Lie symmetries plays a major role in it.

\strut\hfill

	The purpose of this work is to study the symmetry and integrability of all the members of sequence of GMCH equation \cite{GaoLiuQu} is given by
 \begin{eqnarray}
u_{t}-u_{xxt}+2nu_{x}(u^2-u_{x}^2)^{n-1}(u-u_{xx})^2+(u^2-u_{x}^2)^{n}(u_{x}-u_{xxx})=0,
\label{Gmcha}
\end{eqnarray}
When $n=1$ the above equation gives the first member of GMCH and it is also exactly the modified Camassa-Holm Equation (MCH). This MCH equation has been given much attention and it has been analyzed by researchers thoroughly \cite{Fuch-1996, Olver-1996, Qiao-2006, Himonas-2014, Yang-2017}. Also the MCH equation has been discussed in the point of view of symmetries and integrability \cite{Krishnakumar 19}. Interestingly, all the other higher members of the sequence are having dynamically increasing nonlinear terms based on $n$ values ({\it ie}) the nonlinear terms are increasing proportionally to each value of $n$. Therefore, these higher members of the sequence can also be given considerable attention due to the common beautiful properties on symmetries and Painlev\'e Test.

\strut\hfill

In this paper, first we examine the Lie point symmetries and reductions of order for (\ref{Gmcha}) then we prove that the resultant equation is linearizable.\footnotemark[1]{\footnotetext[1]{Throughout this literature the Mathematica add-on Sym \cite{Dimas05a, Dimas06a, Dimas08a} is used to compute the symmetries.}} Next, we apply the Painlev\'e Test for equation (\ref{Gmcha}) and we able to identify that the equation (\ref{Gmcha}) is integrable.

\section {Algebraic properties of all the members of sequence of GMCH}

The generalized modified Camassa-Holm equation is
\begin {equation}
u_{t}-u_{xxt}+2nu_{x}(u^2-u_{x}^2)^{n-1}(u-u_{xx})^2+(u^2-u_{x}^2)^{n}(u_{x}-u_{xxx})=0\label {gmch}
\end {equation}
The first few members of the family are given by
\begin {eqnarray}
u_{t}-u_{xxt}+2u_{x}(u-u_{xx})^2+(u^2-u_{x}^2)(u_{x}-u_{xxx})=0,\label{1}\\
u_{t}-u_{xxt}+4u_{x}(u^2-u_{x}^2)(u-u_{xx})^2+(u^2-u_{x}^2)^{2}(u_{x}-u_{xxx})=0,\label{2}\\
u_{t}-u_{xxt}+6u_{x}(u^2-u_{x}^2)^{2}(u-u_{xx})^2+(u^2-u_{x}^2)^{3}(u_{x}-u_{xxx})=0\label{3}.
\end {eqnarray}
here we can observe that the nonlinearity is increasing as $n$ increases.
The symmetries of the general equation (\ref{gmch}) are
\begin {eqnarray}
& & \textbf{X}_{1} = \partial_{t}, \\
& & \textbf{X}_{2} = \partial_{x}, \\
& & \textbf{X}_{3} =  -2nt \partial t+u \partial u \label{4}.
\end {eqnarray}
Therefore for $n=1,2,3,...$ in (\ref{4}) we have the symmetries for equations $(\ref{1}), (\ref{2}), (\ref{3})...$ respectively. Though the family contains many nonlinear equations, they have similar form of symmetries.
The Lie Brackets of the Lie symmetries of general equation (\ref{gmch}) is\\
\begin{center}
\begin{tabular}{|c|ccc|}
\hline
$\left[ \textbf{X}_{I},\textbf{X}_{J}\right] $ & $\textbf{X}_{1}$ & $\textbf{X}_{2}$ & $\textbf{X}_{3}$  \\
\hline
$\textbf{X}_{1}$ & $0$ & $0$ & $0$   \\
$\textbf{X}_{2}$ & $0$ & $0$ & $2\textbf{X}_{2}$ \\
$\textbf{X}_{3}$ & $0$ & $-2\textbf{X}_{2}$ & $0$ \\
\hline
\end{tabular}
\end{center}
with the algebra $A_1\oplus A_2$.

\strut\hfill

By using the linear combination of $\textbf{X}_{1}$ and $\textbf{X}_{2}$, the traveling wave solution of equation  (\ref{gmch}) can be derived as follows.  The linear combination of the two symmetries $\textbf{X}_{1}$ and $\textbf{X}_{2}$ can be represented as $\textbf{X}_{4}=\partial_{t}+c \partial_{x}$.  The corresponding canonical variables are $r=x-ct$ and $u(x,t)=P(r)$.  Through these canonical variables one can reduce the equation ($\ref{gmch})$ into the following form
\begin{eqnarray}
 c (P^{\prime\prime\prime}-P^{\prime})+(P^2-{P^{\prime}}^2)^n(P^{\prime}
 -P^{\prime\prime\prime})+2nP^{\prime}(P^2-{P^{\prime}}^2)^{n-1}
 (P-P^{\prime\prime})^2=0,  \label{gred2}
\end{eqnarray}
where  $P$ is the function of the new independent variable $r$.  The trivial  symmetry of autonomous differential equation (\ref{gred2}) is $\partial_{r}.$ This symmetry enable us to take the canonical variables $P(r)=p$ and $P^{\prime}=W(p)$.  By using these canonical variables the equation (\ref{gred2}), can be reduced to
\begin{eqnarray}
((p^2-W^2)^{n}-c)(1-{W^{\prime}}^2-WW^{\prime\prime})+
\nonumber\\(p^2-W^2)^{n-1}(2np^2-4npWW^{\prime}+2nW^2{W^{\prime}}^2)=0,         \label{gredsecond}
\end{eqnarray}
where $W$ is a function of $p$.  When we find the symmetries of the equation, ({\ref{gredsecond}}), we arrive the following two possible cases.

\strut\hfill

{\bf Case:1}\\

 If $n\neq-1$, Then the symmetries of the equation ({\ref{gredsecond}})\footnotemark[2]{\footnotetext[2]{These symmetries can also be obtained for all $n\in \mathbb{R}-\{-1\}$, $\mathbb{R}-$denotes the real number}}, are
\begin{eqnarray}
\Gamma_{1}&=&\frac{1}{W((p^2-W^2)^{n}-c)}\partial_{W}\\
\Gamma_{2}&=&\frac{p}{W((p^2-W^2)^{n}-c)}\partial_{W}\\
\Gamma_{3}&=&\partial_{p}+\frac{p(p^2-W^2)^{n}}{W((p^2-W^2)^{n}-c)}\partial_{W}\\
\Gamma_{4}&=&\frac{(p^2-W^2)(c(n+1)-(p^2-W^2)^n)}{(n+1)W((p^2-W^2)^{n}-c)}\partial_{W}\\
\Gamma_{5}&=&4p\partial_{p}\nonumber\\&&-\frac{c(n+1)(3p^2+W^2)-(p^2-W^2)^{n}((4n+3)p^2+W^2)}{(n+1)W((p^2-W^2)^{n}-c)}\partial_{W}\\
\Gamma_{6}&=&2p^2\partial_{p}\\&&-\frac{p(c(n+1)(3p^2+W^2)-(p^2-W^2)^{n}((4n+3)p^2+W^2))}{(n+1)W((p^2-W^2)^{n}-c)}\partial_{W}\nonumber\\
\Gamma_{7}&=&-2\frac{(p^2-W^2)^{n+1}+(n+1)cW^2}{n+1}\partial_{p}\nonumber\\&&-\left(\frac{c^2(n+1)(p^2-3W^2)+2(p^2-W^2)^{2n+1}}
{(n+1)W((p^2-W^2)^{n}-c)}\right.\\&&-\left.\frac{c(p^2-W^2)^{n}(3p^2-(2n+5)W^2)}
{(n+1)W((p^2-W^2)^{n}-c)}\right)p\partial_{W}\nonumber
\end{eqnarray}
\begin{eqnarray}
\Gamma_{8}&=&\frac{2(p^2-W^2)(c(n+1)-(p^2-W^2)^n)}{n+1}p\partial_{p}\nonumber\\&&-\left(\frac{c(n+1)(p^4-W^4)(c(n+1)-(p^2-W^2)^n)}
{(n+1)^2W((p^2-W^2)^{n}-c)}\right.\\&&-\left.\frac{(c(n+1)-(p^2-W^2)^n)(p^2-W^2)^{n+1}((2n+1)p^2+W^2)}{(n+1)^2W((p^2-W^2)^{n}-c)}\right)\partial_{W}\nonumber
\end{eqnarray}
When we use $\Gamma_{3}$, the solution of equation ({\ref{gredsecond}}), is given by
\begin{equation}
\frac{1}{2}cW^2+\frac{(p^2-W^2)^{n+1}}{2(n+1)}+I_{1}=0,
\end{equation}
where $I_{1}$ is the constant of integration.
From $\Gamma_{4}$, the solutions are
\begin{eqnarray}
W&=&\pm p \nonumber\\
W&=&\pm \sqrt{p^2-(c(n+1))^\frac{1}{n}}
\end{eqnarray}
By using $\Gamma_{6}$, the solution is
\begin{equation}
\frac{(p^2-W^2)((p^2-W^2)^{n}-c(n+1))}{p}+I_{2}=0,
\end{equation}
where $I_{2}$ is the constant of integration.
By using $\Gamma_{7}$, the solution is
\begin{equation}
\frac{(n+1)c(p^2-2W^2)-2(p^2-W^2)^{n+1}}{2(n+1)(p^2-W^2)^2(c(n+1)-(p^2-W^2)^{n})^{2}}-I_{3}=0,
\end{equation}
where $I_{3}$ is the constant of integration.

\strut\hfill

{\bf Case:2}\\

If $n=-1$, then the symmetries of the equation ({\ref{gredsecond}}) are
\begin{eqnarray}
\Gamma_{1}&=&\frac{p^2-W^2}{W(c(p^2-W^2)-1)}\partial_{W}\\
\Gamma_{2}&=&\frac{p(p^2-W^2)}{W(c(p^2-W^2)-1)}\partial_{W}\\
\Gamma_{3}&=&\frac{p^2-W^2}{W(c(p^2-W^2)-1)}\partial_{p}+\frac{p(p^2-W^2)}{W(c(p^2-W^2)-1)}\partial_{W}
\end{eqnarray}
\begin{eqnarray}
\Gamma_{4}&=&\frac{(p^2-W^2)(\log{[p^2-W^2]}-c(p^2-W^2)}{W(c(p^2-W^2)-1)}\partial_{W}\\
\Gamma_{5}&=&2p^2\partial_{p}\\&&+\frac{p(p^2-W^2)\log{[p^2-W^2]}-p(2p^2-c(p^4-W^4))}{W(c(p^2-W^2)-1)}\partial_{W}\nonumber\\
\Gamma_{6}&=&4p\partial_{p}\\&&+\frac{(p^2-W^2)\log{[p^2-W^2]}+c(3p^4-W^4)-2p^2(2+cW^2)}{W(c(p^2-W^2)-1)}\partial_{W}\nonumber\\
\Gamma_{7}&=&(\log{[p^2-W^2]}-c(p^2-W^2))\partial_{p}\\&&+\frac{p(c(p^2-W^2)-1)\log{[p^2-W^2]}-p(c^2(p^2-W^2)^2)}{W(c(p^2-W^2)-1)}\partial_{W}\nonumber\\
\Gamma_{8}&=&2p(\log{[p^2-W^2]}-c(p^2-W^2))\partial_{p}\nonumber\\&&+\left(\frac{2W^2(c(p^2-W^2)-1)\log{[p^2-W^2]}}{W(c(p^2-W^2)-1)}\right.\nonumber\\&&
+\left.\frac{c(p^2-W^2)(2p^2-c(p^4-W^4))}{W(c(p^2-W^2)-1)}\right)\partial_{W}\\.
\nonumber
\end{eqnarray}
When we use $\Gamma_{1}$ or $\Gamma_{2}$, the solution of equation ({\ref{gredsecond}}) is
\begin{equation}
W=\pm p.
\end{equation}
From $\Gamma_{3}$, the solutions are
\begin{eqnarray}
W&=&\pm p \nonumber\\
W&=&\frac{p^2}{2}+I_{4}
\end{eqnarray}
By using $\Gamma_{5}$
\begin{equation}
\frac{p}{c(p^2-W^2)-\log{[p^2-W^2]}}+I_{5}=0
\end{equation}
By using $\Gamma_{7}$, the solution is
\begin{equation}
\frac{1}{4}(2\log{[p^2-W^2]}(c(p^2-W^2)-1)-c^2(p^4+W^4)-2cp^2(1-cW^2))+I_{6}=0.
\end{equation}
In the above $I_{1}$, $I_2$, $I_3$, $I_4$, $I_{5}$ and $I_6$ are the constants of integration.

\section{Painlev\'e Property}

We are analyzing the behavior of the singularities of the resultant differential equation by using singularity analysis. The first Sofia Kovalevskaya (1889) \cite{Kovalevskaya 89a, Kovalevskaya 89b, Cooke 84} analyzed the singularity structures for a set of six first-order coupled nonlinear integrable ODEs which is dynamically described the influence of gravity on a heavy rigid body. The singularities are exhibited by the general solution in terms of meromorphic (Jacobian elliptic) functions. Based on Kovalevskaya work, the Painlev\'{e} Property was proposed for ODE's \cite{Painleve97, Painleve00, Painleve02, Hille 76, Ince 56}. If all the movable singularities are single valued then the system of ordinary differential equations said to possess the Painlev\'e Property. The systems are expected to be integrable when they possess Painlev\'e Property. Many researchers are still using Painlev\'e Property to identify new higher-order integrable systems.

\strut\hfill

Almost all the new integrable equations has been already discussed on second-order by considering the general second-order ODE of the form
  \begin{equation}
  \dfrac{d ^2v}{ dr^2} = F \left(r,v,\dfrac{dv}{dr}\right),
  \end{equation}
  where $F$ is rational in $v$, algebraic in $\dfrac{dv}{ dr}$ , locally analytic in $r$. Especially, there were only fifty equations possessing the Painlev\'e Property. The solutions which is in terms of elementary functions including the elliptic functions can be obtained only for forty four differential equations out of the fifty integrable equations. The remaining six are referred as Painlev\'e transcendental equations \cite{Ince 56}. Recently both Right and Left Painlev\'e Series constitutes the solution of the given differential equation within a punctured disc centred on the singularity has studied in \cite{Feix 97 a, Andriopoulos 06 a, Lemmer 93 a, Krishnakumar 16}.

\strut\hfill

Ablowitz {\it et al.} have given an algorithmic procedure which is called Ablowitz-Ramani-Segur (ARS) algorithm \cite{Ablowitz 80a, Ablowitz 80b, Ramani 89} to perform the Painlev\'e Test. This algorithm has the following three steps

 \begin{enumerate}
   \item Leading-order behavior.
   \item Determination of resonances.
   \item Arbitrariness of sufficient numbers of constants.
 \end{enumerate}

\subsection{Painlev\'e Test}

 We are very much interested to apply Painlev\'e Test for a third-order resultant ordinary differential equation which is obtained from the following partial differential equation (\ref{gmch})
\begin{equation}
u_{t}-u_{xxt}+2nu_{x}(u^2-u_{x}^2)^{n-1}(u-u_{xx})^2+(u^2-u_{x}^2)^{n}(u_{x}-u_{xxx})=0.\label{gmch1}
\end {equation}

  The following resultant third-order nonlinear autonomous ordinary differential equation which obtain from to equation (\ref{gmch1}) by substituting $r=x-c t$ is given by

\begin{eqnarray}
 c( P^{\prime\prime\prime}-P^{\prime})+(P^2-{P^{\prime}}^2)^n(P^{\prime}
 -P^{\prime\prime\prime})+2nP^{\prime}(P^2-{P^{\prime}}^2)^{n-1}
 (P-P^{\prime\prime})^2=0,  \label{Ggred2}
\end{eqnarray}
where $P$ is a function of $r$. This equation (\ref{Ggred2}) does not pass the Painlev\'e Test. But, a substitution, $P(r)=\dfrac{1}{v[r]},$ for $P(r)$ shall make the above resultant equation as integrable. The resultant equation is given by \footnotemark[3]{\footnotetext[3]{The resultant equations of all members of the sequence can be obtained from (\ref{Oeqn}) by taking the values of $n=1,2,3,...$}}
 \begin{eqnarray}\label{Oeqn}
 6 c ^{4n}{v^{\prime}}^5-2(4n+3){v^{\prime}}^5\sum_{k=0}^n(-1)^k{n \choose k}v^{2n-2k}{v^{\prime}}^{2k}+6 c v^{4n+3}v^{\prime}v^{\prime\prime}-\nonumber\\6 c v^{4n+1}{v^{\prime}}^3v^{\prime\prime}
 -2(2n+3)v^3v^{\prime}v^{\prime\prime}\sum_{k=0}^n(-1)^k{n \choose k}v^{2n-2k}{v^{\prime}}^{2k}+\nonumber\\2(4n+3)v {v^{\prime}}^3v^{\prime\prime}\sum_{k=0}^n(-1)^k{n \choose k}v^{2n-2k}{v^{\prime}}^{2k}+c v^{4n+4}(v^{\prime}-v^{\prime\prime\prime})-\nonumber\\c v^{4n+2}{v^{\prime}}^2(7v^{\prime}-v^{\prime\prime\prime})-v^4((2n+1)
 v^{\prime}-v^{\prime\prime\prime})\sum_{k=0}^n(-1)^k{n \choose k}v^{2n-2k}{v^{\prime}}^{2k}+\nonumber\\ v^2v^{\prime}((8n+7){v^{\prime}}^2-
 2n{v^{\prime\prime}}^2-v^{\prime}v^{\prime\prime\prime})\sum_{k=0}^n(-1)^k{n \choose k}v^{2n-2k}{v^{\prime}}^{2k}=0.
 \end{eqnarray}
 According to the Painlev\'e Test the dominant terms of (\ref{Oeqn}) are given by
\begin{eqnarray}\label{DOeqn}
6 c v^{4n}{v^{\prime}}^5-(-1)^n 2(4n+3){v^{\prime}}^5{v^{\prime}}^{2n}-
6cv^{4n+1}{v^{\prime}}^3v^{\prime\prime}
+\nonumber\\(-1)^n 2 (4n+3)v {v^{\prime}}^{2n+3}v^{\prime\prime}+
c v^{4n+2}{v^{\prime}}^2v^{\prime\prime\prime}-
\nonumber\\(-1)^n v^2{v^{\prime}}^{2n+1}(2n{v^{\prime\prime}}^2-v^{\prime}v^{\prime\prime\prime})
\end{eqnarray}
Here, $v=a_{-1} w^{-1}$, where $w=(r-r_0),$ is a leading-order behavior of the Laurent series in the neighborhood of the movable singular point $r_0$. To determine the resonances we substitute $v(w)=a_{-1} w^{-1} + m w^{-1 + s}$ into (\ref{DOeqn}) and by equating the coefficients of $m$ to zero then we have $s=-1, \, 0 ~\text{and} ~1$. Therefore, the resonances show that the powers zero and one of $w$ at which arbitrary constants of the solution for (\ref{Oeqn}) can enter into the Laurent series expansion. In this series we have sufficient number of arbitrary constants. Hence, the equation (\ref{gmch1}) is integrable as consequence of Painlev\'e property holds for equation (\ref{Ggred2}).

\section {Conclusion}

 Even though the nonlinearity increasing as $n$ increases, all members of the family which is obtained from the generalized modified Camassa-Holm (GMCH) Equation have similar form of symmetries, reductions and solutions. We have obtained two sets of eight dimensional symmetries ({\it ie} maximum number of symmetries) for resultant general second-order differential equation (\ref{gredsecond}) of the family for $n=-1$ and $n\in \mathbb{R}-\{-1\}$ respectively. Therefore all members of the family of second-order differential equation are linearizable.

 \strut\hfill

 By the performance of Painlev\'e Test for the resultant generalized third-order differential equation (\ref{Ggred2}) we have observed that all members of the sequence of generalized modified Camassa-Holm (GMCH) Equation posses Painlev\'e Property. In particular, leading order, resonances and arbitrariness of constant of solutions are also having similar form for the entire members of the sequence. Hence all members of the sequence of generalized modified Camassa-Holm (GMCH) Equation are integrable.

\section*{Acknowledgements}
KK and ADD thank Prof. Stylianos Dimas, S\'ao Jos\'e dos Campos/SP, Brasil, for providing a new version of the SYM-Package. KK thank Late Prof. K.M.Tamizhmani for his support and academic guidance during his Doctoral programme.

\end{document}